%% file: fuzzwiz.tex
\documentclass[conference, letterpaper, twocolumn]{IEEEtran}
\IEEEoverridecommandlockouts
\usepackage[colorlinks=false]{hyperref}
\usepackage{amsmath,amssymb,amsfonts}
\usepackage{algorithmic}
\usepackage{graphicx}
\usepackage{textcomp}
\usepackage{enumitem}
\usepackage[utf8]{inputenc}
\usepackage[T1]{fontenc} 
\usepackage[nolist]{acronym}
\usepackage[english]{babel}
\usepackage{csquotes}
\usepackage{amsmath}											% ams stuff
\usepackage{amsfonts}											% ...
\usepackage{blindtext} 											% most important package when writing papers :)
\usepackage{booktabs}											% toprule, midrule, bottomrule
\usepackage{fancyvrb}
\usepackage{ifthen}												% for 'if-then-else' statements
\usepackage{filecontents} 										% file import

\usepackage{graphicx}
\usepackage[dvipsnames,table]{xcolor}
\usepackage{ifthen}												% for 'if-then-else' statements
\usepackage{listings}											% Listings
\usepackage{lipsum}
\usepackage{multirow}
\usepackage{pgfplots}
\usepackage{siunitx}
\usepackage{ragged2e}						
\usepackage{tabularx}
\usepackage{url}
\usepackage{tikz}
\usepackage{circuitikz}
\usepackage{comment}
\usepackage{footnote}
\usepackage{float}
\usepackage{listings}
\usepackage{setspace}
\usepackage{makecell, tabularx}

\usetikzlibrary{calc,trees,positioning,arrows,chains,shapes.geometric,%
	decorations.pathreplacing,decorations.pathmorphing,shapes,%
	matrix,shapes.symbols}

\usepackage[font=footnotesize,labelfont=bf]{subcaption}		
\usepackage{caption}
\usepackage[ruled,linesnumbered]{algorithm2e}
\SetAlCapNameFnt{\small}
\SetAlCapFnt{\small}
\SetAlgoNlRelativeSize{0}
\RestyleAlgo{ruled}
\SetKwComment{Comment}{/* }{ */}
\usepackage[style=ieee, backend=biber, natbib=true, maxbibnames=1, minbibnames=1]{biblatex}
\usepackage{fancyhdr}
\fancypagestyle{firstpage}{%
	\fancyhf{}%
	\fancyhead[C]{\normalsize To appear at the 16th International Symposium on Electronics and Telecommunications (ISETC'24), Nov 7-8, 2024, Timisoara, Romania}
	
	\fancyfoot[C]{\footnotesize \copyright2024 IEEE. Personal use of this material is permitted. Permission from IEEE must be obtained for all other uses, in any current or future media, including reprinting/republishing this material for advertising or promotional purposes, creating new collective works, for resale or redistribution to servers or lists, or reuse of any copyrighted component of this work in other works.}
}

\newboolean{blindreview}
\setboolean{blindreview}{false}
\DeclareRobustCommand{\IEEEauthorrefmark}[1]{\smash{\textsuperscript{\footnotesize #1}}}
\usepackage[a4paper, total={184mm,239mm}]{geometry}
\begin{document}	
	
%%%%%%%%%%%%%%%%%%%%%%%%%%%%%%%%%%%%%%%%%%%%%%%%%%%%%%%%%%%%%%%%%%%%%%%%%%%%%%%%%%%%%%%%
%%%%% Acronyms
%%%%%%%%%%%%%%%%%%%%%%%%%%%%%%%%%%%%%%%%%%%%%%%%%%%%%%%%%%%%%%%%%%%%%%%%%%%%%%%%%%%%%%%%
\acrodefplural{ADC}[ADC]{Analog-to-Digital Converters}
\acrodefplural{SoCs}{System-on-Chips}
%\acrodefplural{HTs}{Hardware Trojans}
\acrodefplural{IC}[IC]{Integrated Circuits}
\acrodefplural{IP}[IP]{Intellectual Properties}
\acrodefplural{HDLs}{Hardware Description Languages}
\acrodefplural{CSRs}{Control and Status Registers}
\begin{acronym}[placeholder]
	
	\acro{SoC}{System-on-Chip}
	\acro{CRV}{Constrained Random Verification}
	\acro{CDG}{Coverage Directed Test Generation}
	\acro{ASIC}{Application-Specific Integrated Circuit}
	\acro{IP}{Intellectual Property}
	\acro{HT}{Hardware Trojan}
	\acro{RL}{Reinforcement Learning}
	\acro{RTL}{Register Transfer Level}
	\acro{IC}{Integrated Circuit}	
	\acro{DUV}{Design Under Verification}
	\acro{DUT}{Design Under Test}
	\acro{EDA}{Electronic Design Automation}
	\acro{HDL}{Hardware Description Language}
	\acro{HSB}{Hardware Simulation Binary}
	\acro{AFL}{American Fuzzy Lop}
	\acro{ISA}{Instruction Set Architecture}
	\acro{CSR}{Control and Status Register}
	\acro{MDA}{Model Driven Architecture}
	\acro{UML}{Unified Modeling Language}
	\acro{NRE}{Non-Recurring Engineering}
	\acro{API}{Application Programming Interface}
	\acro{GUI}{Graphical User Interface}
	\acro{XML}{Extensible Markup Language}
	\acro{ToM}{Template of MetaFuzz}
	\acro{CSV}{Comma Separated Values}
	\acro{AES}{Advanced Encryption Standard}
	\acro{KMAC}{Keccak Message Authentication Code}
	\acro{HMAC}{Hash-based Message Authentication Code}
	\acro{RV-Timer}{RISC-V Timer}
	\acro{IPs}{Intellectual Properties}
	\acro{ToTB}{Template of Testbench}
	\acro{CWE}{Common Weakness Enumeration}
\end{acronym}
%%%%%%%%%%%%%%%%%%%%%%%%%%%%%%%%%%%%%%%%%%%%%%%%%%%%%%%%%%%%%%%%%%%%%%%%%%%%%%%%%%%%%%%%
%%%%%%%%%%%%%%%%%%%%%%%%%%%%%%%%%%%%%%%%%%%%%%%%%%%%%%%%%%%%%%%%%%%%%%%%%%%%%%%%%%%%%%%%

\title{FuzzWiz - Fuzzing Framework for Efficient Hardware Coverage \vspace{-0.3cm}

\thanks{This work has been developed in the project VE-VIDES (project label 16ME0243K) which is partly funded within the Research Programme ICT 2020 by the German Federal Ministry of Education and Research (BMBF)}
}

\ifthenelse{\boolean{blindreview}}{}{
	\author{\IEEEauthorblockN{
			Deepak Narayan Gadde\IEEEauthorrefmark{1},
			Aman Kumar\IEEEauthorrefmark{1},
			Djones Lettnin\IEEEauthorrefmark{2},
			Sebastian Simon\IEEEauthorrefmark{1}}
		\IEEEauthorblockA{
			\IEEEauthorrefmark{1}Infineon Technologies Dresden GmbH \& Co. KG, Germany \\
			\IEEEauthorrefmark{2}Infineon Technologies AG, Germany}
	\vspace{-1.1cm}}   
}

\maketitle
\thispagestyle{firstpage}
%\vspace*{-1cm}

\begin{abstract}

Ever-increasing design complexity of \acp{SoC} led to significant verification challenges. Unlike software, bugs in hardware design are vigorous and eternal i.e., once the hardware is fabricated, it cannot be repaired with any patch. Despite being one of the powerful techniques used in verification, the dynamic random approach cannot give confidence to complex \ac{RTL} designs during the pre-silicon design phase. In particular, achieving coverage targets and exposing bugs is a complicated task with random simulations. In this paper, we leverage an existing testing solution available in the software world known as \textit{fuzzing} and apply it to hardware verification in order to achieve coverage targets in quick time. We created an automated hardware fuzzing framework \textit{FuzzWiz} using metamodeling and Python to achieve coverage goals faster. It includes parsing the RTL design module, converting it into C/C++ models, creating generic testbench with assertions, fuzzer-specific compilation, linking, and fuzzing. Furthermore, it is configurable and provides the debug flow if any crash is detected during the fuzzing process. The proposed framework is applied on four IP blocks from Google's OpenTitan chip with various fuzzing engines to show its scalability and compatibility. Our benchmarking results show that we could achieve around \SI{90}{\percent} of the coverage 10 times faster than traditional simulation regression based approach.

\end{abstract}

%This problem could not be solved by methodologies such as constrained random verificaton used in coverage guided test generation.
%Our experiments showed that choosing the right engine for fuzzing process is vital to reach higher hardware line coverage.
%However, Fairfuzz performed better than AFL++ for two of those four IPs and marginally improved the coverage


\section{Introduction} \label{sec:introduction}

\input{sections/introduction}

\section{Background} \label{sec:background}
\input{sections/background}

\section{Related Work} \label{sec:related_work}

\input{sections/related_work}

\section{Proposed Fuzzing Framework} \label{sec:methodology}
\input{sections/methodology}

\section{Results and Discussion} \label{sec:results}

\input{sections/results}

\section{Conclusion} \label{sec:conclusion}
\input{sections/conclusion}

\vspace{0.25cm}

\printbibliography 

\end{document}

%% file: sections/introduction.tex
Design verification has become the bottleneck with the increasing complexity of \acp{SoC}, accounting for more than \SI{60}{\percent} of the total project time \cite{VerStudy}. The simulation-based verification is the strongest method in verification and it entails simulating a \ac{DUV} with valid input sequences and evaluating the \ac{DUV}'s behavior during or after simulation to discover bugs in the design \cite{asic_soc}. \ac{CRV} is one of the most widely used methodologies, it creates test scenarios by randomizing input sequences. Although successful in identifying \ac{RTL} design bugs, this method falls short in thoroughly exploring the state space of the design due to the phenomenon of time-space explosion. Additionally, \ac{CDG} is another renowned dynamic method which is deployed in both software testing and hardware verification. In this technique, coverage metrics from the present simulation are used to drive the generation of inputs for subsequent simulations. Over the last two decades, numerous solutions have been proposed for \ac{CDG}. However, effectiveness of the test generation depends on the structure of the \ac{DUV}, the coverage criterion, and input space. The generated tests still miss a large number of potentially severe bugs \cite{3384159}. Regression of these test simulations often require longer turnaround time to reach sensible coverage metrics \cite{deepak_ml}. Hence, researchers are still in search for an efficient solution to reduce these turnaround times and to improve verification throughput. 

Few of those solutions include the \textit{coverage-guided fuzzing} technique which is used widely in software testing for security assessment and to achieve efficient coverage of the program under test. Several studies on fuzzing in hardware verification have been conducted in recent years to address the \ac{CDG} problem \cite{3638046}. These studies, while yielding positive outcomes, have certain limitations in terms of \ac{HDL} used for RTL implementation, type of coverage feedback, and the fuzzing engine used (Sec. \ref{sec:related_work}). This caught our attention and motivated us to create an automated framework to perform hardware fuzzing.

\begin{comment}
In this technique, the fuzzer generates and executes test cases with the primary goal of maximizing code coverage in the target. A fuzzer accomplishes this by continuously monitoring which paths or functions have been exercised and then focusing its efforts to investigate untested or under-tested areas. As it utilizes coverage metrics to drive the creation of additional inputs or stimuli, the method is called coverage-directed.
\end{comment}

Our contributions to this work are sketched as follows:
\begin{itemize}
	\item Automated fuzzing framework using \textit{Python} and metamodel - \textit{FuzzWiz} (Sec. \ref{sec:methodology})
	\item Metamodeling code generation framework to produce fuzzer specific testbench - \textit{MetaFuzz} (Sec. \ref{sec:methodology_metafuzz})
	\item Benchmarking different fuzzing engines based on their performance on OpenTitan IP cores \cite{opentitan} in terms of coverage (Sec. \ref{sec:results})
\end{itemize}

%% file: sections/background.tex
In this section, we provide the basis for the fuzzing technique and corresponding tools used for it i.e., fuzzers together with an introduction to the metamodel framework. 

\subsection{Coverage-guided Fuzzing}

\begin{figure}[htbp]
	\centering
	\includegraphics[width=0.7\linewidth]{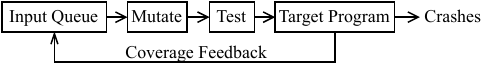}
	\caption{Coverage-guided fuzzing}
	\label{fuzzing_picture}
\end{figure}
\vspace{-0.3cm}

The fundamental process of coverage-guided fuzzing is shown in Fig. \ref{fuzzing_picture} which includes input generation and mutation, test execution, error detection, analysis, and a feedback loop \cite{gen_fuzzing}. These steps repeat in a systematic manner to uncover vulnerabilities in the target system by covering most of the input combinations. It continuously monitors and analyzes the code coverage achieved during testing, prioritizing inputs that lead to unexplored or less-covered code paths within the target application. By doing so, it maximizes the chances of identifying critical security flaws, crashes, or unexpected behaviors \cite{9780992}. This feedback-driven methodology has proven to be most valuable and has helped in achieving the target coverage and detecting crashes in software \cite{8233151}.

Fuzzing could be applied to hardware either by translating hardware to a software model or by directly fuzzing hardware. According to prior works \cite{google_interns}, \cite{9256500}, \cite{tyagi2022thehuzz}, \cite{8587711}, both approaches have shown promising results but have shortcomings. Although fuzzing could be done directly on the hardware, we need to rely on software based metrics to determine the coverage targets. In our work, we convert the hardware into software and fuzz it using various open-source software fuzzing engines those utilize different mutation strategies. The comparison among those engines namely, \textit{AFL} \cite{afl_bib}, \textit{AFL++} \cite{aflpp_bib}, \textit{Fairfuzz} \cite{fairfuzz_bib}, \textit{Perffuzz} \cite{perffuzz_bib}, and \textit{Tortoisefuzz} \cite{tortoisefuzz_bib} is shown in Table \ref{fuzzer_comparison}.

\begin{table}[htbp]
	\setcellgapes{1pt}
	\makegapedcells
	\renewcommand{\arraystretch}{1.5}
	\caption{Comparison between various software fuzzing engines}
	\label{fuzzer_comparison}
	\scriptsize
	\centering
	\resizebox{0.5\textwidth}{!}{
		\begin{tabular}{|c|l|l|l|l|}
			\hline
			\textbf{Name}         & \textbf{Year} & \textbf{Feedback}    & \textbf{Search Strategy}       & \textbf{Mutation Strategy}    \\ \hline
			\textbf{AFL}          & 2013          & Coverage             & Fitness                        & Evolutionary                  \\ \hline
			\textbf{Fairfuzz}     & 2018          & Coverage/ Fairness   & Rare-branch covering inputs    & Evolutionary (Fairness)       \\ \hline
			\textbf{Perffuzz}     & 2018          & Coverage             & Multi-dimensional feedback     & Evolutionary (Havoc)          \\ \hline
			\textbf{AFL++}        & 2019          & Coverage             & Fitness (shortest and fastest) & Evolutionary (Markov's model) \\ \hline
			\textbf{Tortoisefuzz} & 2020          & Coverage  & Cost-favored                   & Genetic algorithms            \\ \hline
		\end{tabular}
	}
\end{table}

\subsection{Metamodeling}
Metamodeling frameworks are automation frameworks to lower non-recurring engineering expenditures such as \textbf{generic code generation} in this work. They make use of \ac{UML} diagrams, Python, and Mako templates \cite{mako_templates} to generate outputs in various languages. Metamodeling is based on the concept of model driven architecture, which emphasizes the use of modeling techniques to boost productivity and improve the level of abstraction during the development of software and hardware systems \cite{8715154}. 
%A model represents a system at a particular abstraction level and a metamodel depicts a model's structure and the relationships between its constituent parts.

\begin{figure}[htbp]
	\centering
	\includegraphics[width=0.5\textwidth]{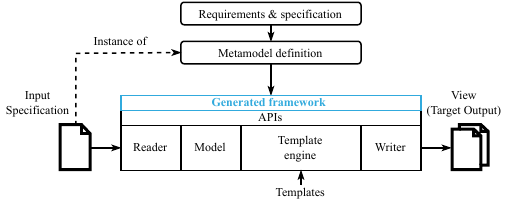}
	\caption{Metamodel-based automation framework}
	\label{metamodel}
\end{figure}

The metamodel-based code generator framework is depicted in Fig. \ref{metamodel}. The initial step of it is to create a \ac{UML} class diagram to define a metamodel, which is then used to generate Python \acp{API} for easy access and manipulation of the metamodel through common methods like \textit{setters()} and \textit{getters()}. These \acp{API} interface with an \ac{XML}-based design specification, allowing a \textit{reader} component to interpret the \ac{XML} and a \textit{writer} to output target code in various languages through the use of Mako templates. The framework's flexibility enables the production of codes for a variety of languages or applications, such as RTL design code, SystemVerilog testbenches, assertions \cite{8644957}, and C++ files, by simply using appropriate template files.

%% file: sections/related_work.tex
All relevant prior works on fuzzing the hardware designs are discussed in this section. RFUZZ \cite{8587711} was an early endeavor that performed coverage-directed mutational fuzzing directly on \ac{HDL}, employing \textit{mux toggle coverage} to steer the process and showing promising results across various \ac{RTL} designs. \cite{google_interns} proposed a design-agnostic technique to convert HDL code into software binaries, enabling the use of the \ac{AFL} fuzzer for hardware design. This method, complemented with generic interfaces and binary instrumentation for tracing hardware coverage in software, achieved substantial HDL code coverage on OpenTitan IP designs. DirectFuzz \cite{9586289} introduced targeted gray-box fuzzing, producing test inputs aimed at specific RTL module instances, and delivered faster coverage compared to RFUZZ. Other works such as DIFUZZRTL \cite{9519470}, focused on \textit{register-coverage} guidance to unearth CPU bugs in RTL designs. And ProcessorFuzz \cite{10133714}, which utilized a novel \textit{\ac{CSR}-transition coverage} metric to monitor processor state changes, proved more efficient in bug detection than previous methods.

Most of the research \cite{8587711}, \cite{9519470}, \cite{10133714} discussed earlier used specific metrics such as \textit{mux toggle}, \textit{CSR-transition}, or \textit{register} coverage, to guide the fuzzing technique and various fuzzing tools. It is questionable how their specific coverage metrics relate to traditional \ac{RTL} code or functional coverage which are needed to ensure that the \ac{DUV} is verified. Moreover, some of the prior works \cite{8587711} could not be applied on any generic \ac{RTL} design. Although one cannot completely guarantee the flawless condition of a system due to the random and unpredictable behavior of fuzzers, their capability for automation provides great potential to be a valuable supplement to the existing verification methods.

In our work, we developed an automation framework which takes the generic \ac{RTL} design(s) as an input and performs fuzzing to improve the overall \ac{HDL} coverage with flexibility to choose which fuzzing engine should be applied. Some of the work from \cite{google_interns} is reused in this work to perform a statistical comparison of our coverage results.

\begin{comment}
\begin{figure*}[htp]
	\centering
	\includegraphics[width=\textwidth]{figures/fuzzframework.pdf}
	\caption{FuzzWiz - An automated fuzzing framework}
	\label{fuzzwiz}
\end{figure*}
\end{comment}

%% file: sections/methodology.tex
%\begin{comment}

%\end{comment}
\begin{figure*}[t!]
	\centering
	\includegraphics[width=0.85\textwidth]{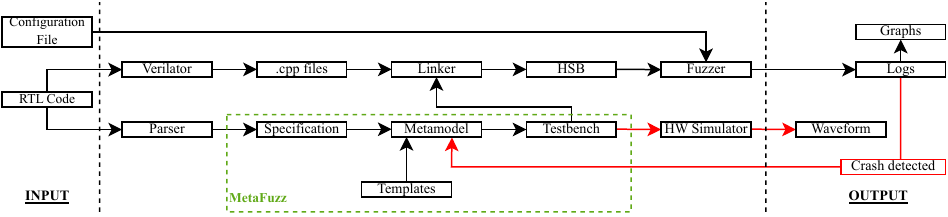}
	\caption{FuzzWiz - Automated hardware fuzzing framework}
	\label{fuzzwiz}
\end{figure*}
This section presents the proposed fuzzing framework \textit{FuzzWiz} as shown in Fig. \ref{fuzzwiz}. The main steps involved in the \textit{FuzzWiz} framework are demonstrated with a simple \ac{RTL} module given in Listing \ref{example_rtl}. This design has a register intended for the secure storage of cryptographic keys with an additional debug feature. Consider a critical security requirement here that the keys must remain inaccessible during debug mode to prevent unauthorized access. 

\begin{lstlisting}[language=verilog,
	numbers=none,
	caption= A simple RTL module (used only for demonstration purpose),
	captionpos=b,
	frame=bt,
	label=example_rtl,
	linewidth=\columnwidth,
	breaklines=true,
	basicstyle=\scriptsize,
	floatplacement=htbp,
	columns=fullflexible]
module key_store_debug (input wire clk_in, input wire rst_n_in, input wire debug_mode, output reg [63:0] key_out);
  reg [31:0] AES_ENC_DEC_KEY_0 = 32'hACEC;
  reg [31:0] AES_ENC_DEC_KEY_1 = 32'hBADE;
  reg [3:0] debug_info;
  reg [63:0] out;
  always @(posedge clk_in or negedge rst_n_in) begin
    if (!rst_n_in) begin
      debug_info = 0;
      out = 0;
    end 
    else if (debug_mode)
      debug_info = debug_info+1;
    out = {AES_ENC_DEC_KEY_1, AES_ENC_DEC_KEY_0};
  end
  assign key_out = out;
endmodule
\end{lstlisting}
\vspace{-0.3cm}

\subsection{Configuration of the framework}
A \textit{hjson} file is used for configuring the \textit{FuzzWiz} framework which contains attributes, the preferred fuzzing engine, and the duration to perform the fuzzing process. These values are updated by the user based on their preferred fuzzing engine and on how long the \ac{RTL} code should be fuzzed.

For demonstration, \textit{FuzzWiz} is configured to run fuzzing on the example design (Listing \ref{example_rtl}) for one hour using the \textit{Fairfuzz} engine to reach efficient hardware coverage.

\subsection{Generation of software models}
Most open-source simulators initially transform the hardware \ac{RTL} into software representations, typically in the form of \textit{C} or \textit{C++} models, prior to conducting simulations. To facilitate fuzzing of hardware designs, this software model is intercepted and used by fuzzer. In our \textit{FuzzWiz} framework, we employ Verilator \cite{verilator} for the conversion of \ac{RTL} files into equivalent \textit{C++} files.

\subsection{Extraction of specification from the RTL}
All the primary ports of \ac{DUV}, along with the other data such as direction of the ports, size and type of ports, and top module name are extracted using an RTL parser. This information is used to create a formalized specification in \textit{XML} format.

\subsection{Creation of generic testbench} \label{sec:methodology_metafuzz}
The formalized specification is transferred to the \ac{ToTB} within the intermediary \textit{MetaFuzz} layer. The \ac{ToTB}, which is based on \textit{Python}, is responsible for the extraction of design details from the specification and the subsequent definition of \textit{MetaFuzz} models for each necessary testbench file. Following this, the view layer of \textit{MetaFuzz} is responsible for mapping these models onto a specific language, such as \textit{SystemVerilog} or \textit{C++}. This mapping facilitates the generation of a generic testbench in \textit{SystemVerilog} and fuzzer-specific files in \textit{C++}, tailored to the given \ac{RTL}. For example, the testbench in Listing \ref{example_tb} is generated using metamodel \textit{MetaFuzz} for the design in Listing \ref{example_rtl}.
\begin{lstlisting}[language=verilog,
	numbers=none,
	caption=Generic testbench generated using metamodel,
	captionpos=b,
	label=example_tb,
	frame=bt,
	linewidth=\columnwidth,
	breaklines=true,
	basicstyle=\scriptsize,
	floatplacement=htbp,
	columns=fullflexible]
	module key_store_debug_tb (input wire clk_in, input wire rst_n_in, input wire debug_mode, output reg [63:0] key_out);
	key_store_debug cl(.clk_in(clk_in),
	.rst_n_in(rst_n_in),
	.debug_mode(debug_mode),
	.key_out(key_out));
	endmodule
\end{lstlisting}
\vspace{-0.3cm}
Two assertions as shown in Listing \ref{example_assertions} are added manually to the testbench to verify the security requirement of the example design. The crashes caused during the fuzzing process are used to demonstrate the RTL debug flow.

\begin{lstlisting}[language=verilog,
	numbers=none,
	caption=Assertions for the given RTL to expose bugs,
	captionpos=b,
	frame=bt,
	label=example_assertions,
	linewidth=\columnwidth,
	breaklines=true,
	basicstyle=\scriptsize,
	floatplacement=htbp,
	columns=fullflexible]
assert property(@(posedge clk_in) disable iff(!rst_n_in) debug_mode== 1 |->key_out == 0);
assert property(@(posedge clk_in) disable iff(!rst_n_in) debug_mode== 0 |->key_out == 64'h0000BADE0000ACEC);
\end{lstlisting}
\vspace{-0.3cm}

\subsection{Generation of a HSB}
Before initiating the fuzzing process, the converted software \textit{C++} models are instrumented using a specific compiler depending on the chosen fuzzing engine. All the five fuzzers used in this work offer their own compiler variant. These specialized compilers perform source code instrumentation to enhance the fuzzing technique. The instrumented files in conjunction with the \textit{SystemVerilog} testbench files are linked to generate a \ac{HSB}.

\subsection{Fuzzing and visualization of coverage output}
The generated \ac{HSB} will be fuzzed by the selected fuzzing engine for a certain duration depending on the framework configuration. After the fuzzing process, coverage details are extracted individually for each testcase generated by the fuzzer. Open-source tools such as kcov \cite{kcov}, LLVM \cite{1281665}, and Verilator \cite{verilator} are integrated in \textit{FuzzWiz} and are used to trace and merge various coverage data such as software line, block, and hardware line coverage respectively. Subsequently, the collected data is used by \textit{Python} automation scripts to create \textit{CSV} files for the specific coverage type and to plot graphs. This paper focuses on results related to hardware line coverage. 

\subsection{RTL debug flow}

\begin{figure}[htbp]
	\centering
	\includegraphics[width=0.6\linewidth]{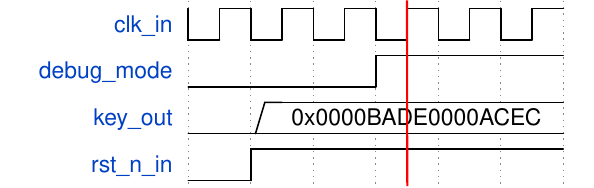}
	\caption{RTL simulation for root cause analysis}
	\label{wavedrom}
\end{figure}

The inputs that resulted in a software crash during the fuzzing process are recorded for \ac{RTL} debugging purposes. The Metamodel \textit{MetaFuzz} generates a \textit{SystemVerilog} testbench containing the input sequence that caused the crash. With the aid of any hardware simulator, this testbench can simulate the \ac{DUV} and provide a waveform that can be utilized for finding the root cause for a software crash at \ac{RTL} level.

For instance, one of the crashes during the fuzzing of the \ac{DUV} (Listing \ref{example_rtl}) is analyzed as follows: After resetting the \ac{DUV}, the testbench receives input stimuli (i.e., \textit{debug\_mode}) from the fuzzer. When \textit{debug\_mode} is \textit{0x1}, the first assertion fails as the stored key \textit{0x0000BADE0000ACEC} is observed at the output pin. This assertion failure is recorded as a crash, the corresponding inputs are also captured. A testbench is generated using metamodeling with these inputs to run a \ac{RTL} simulation as shown in Fig. \ref{wavedrom}

%%%%%%%%%%%%%%%%%%%%%%%%%%%
%%%     RESULTS
%%%%%%%%%%%%%%%%%%%%%%%%%%%

\begin{figure*}[ht!]
	\centering	
	\begin{subfigure}{0.24\textwidth}
		\includegraphics[width=\linewidth]{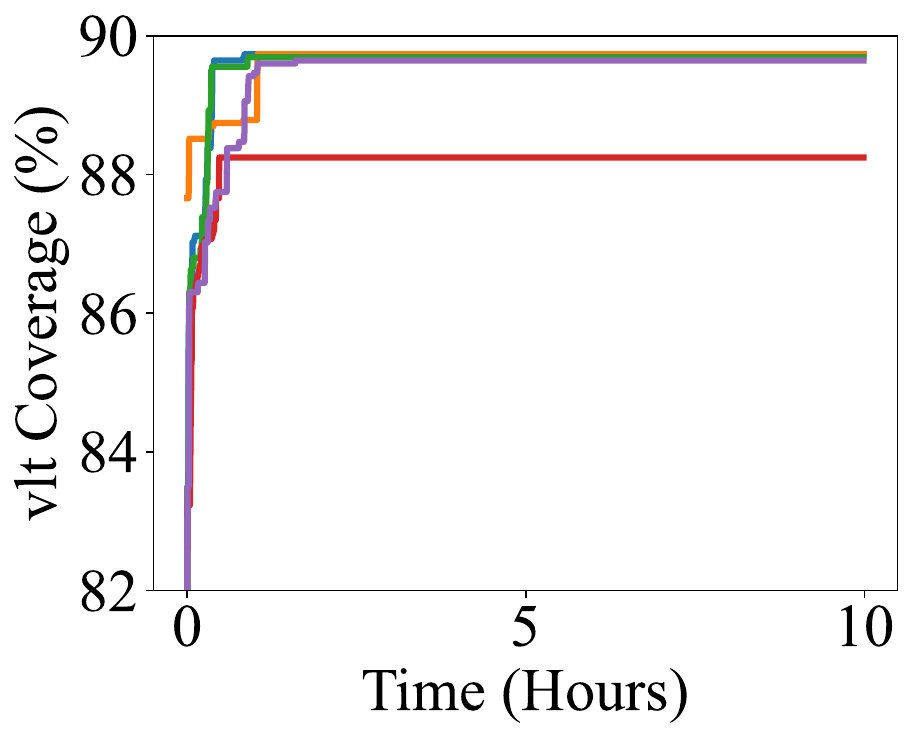}
		\caption{aes | HW Line (vlt)}
		\label{aes_HW_Line}
	\end{subfigure}
	\hfill
	\begin{subfigure}{0.24\textwidth}
		\includegraphics[width=\linewidth]{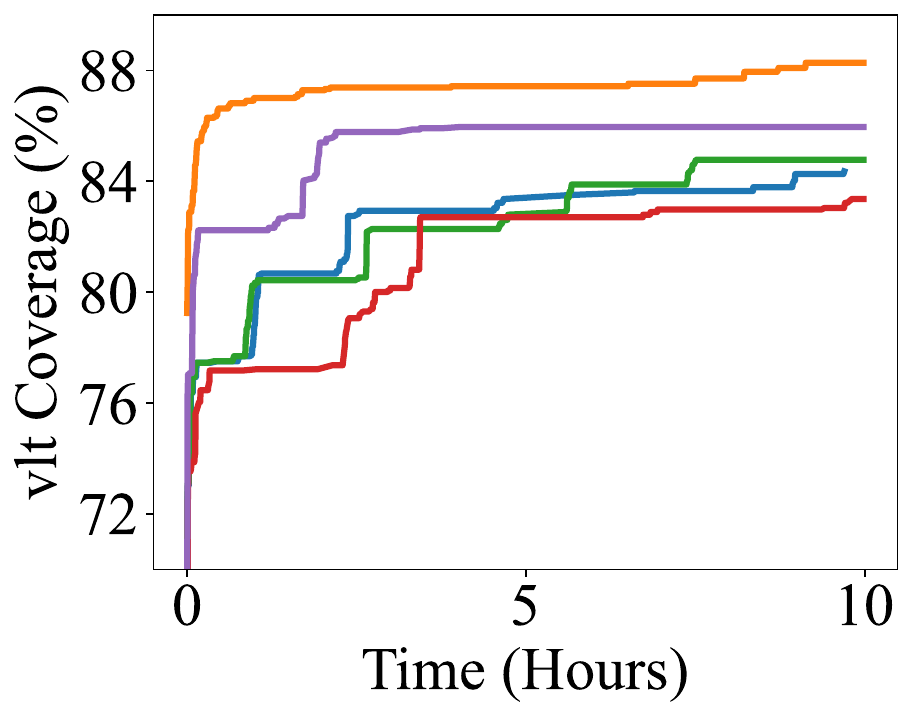}
		\caption{kmac | HW Line (vlt)}
		\label{kmac_HW_Line}
	\end{subfigure}
	\hfill
	\begin{subfigure}{0.24\textwidth}
		\includegraphics[width=\linewidth]{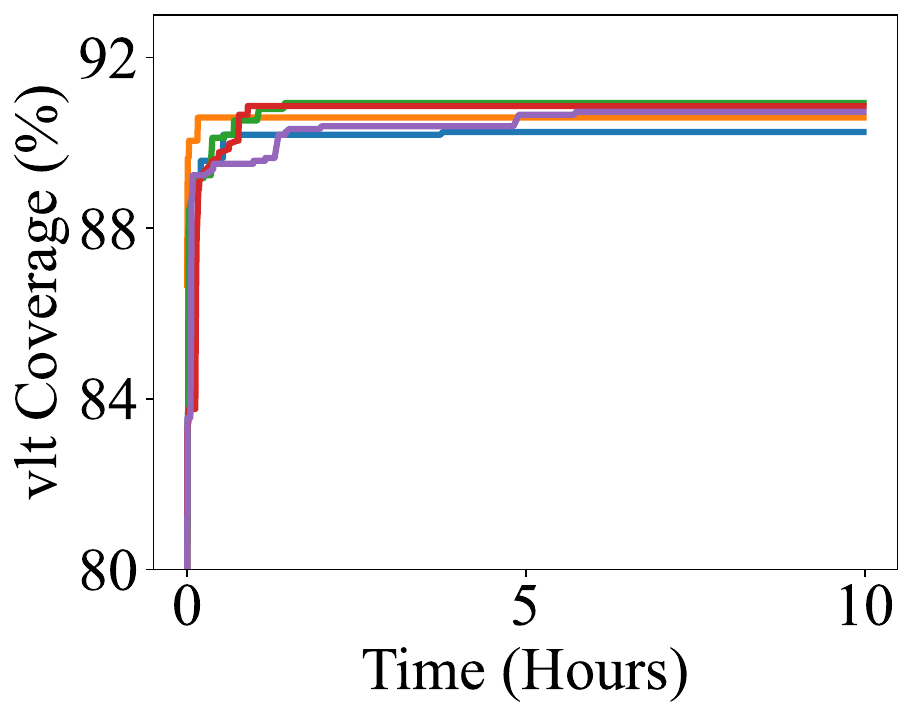}
		\caption{hmac | HW Line (vlt)}
		\label{hmac_HW_Line}
	\end{subfigure}
	\hfill
	\begin{subfigure}{0.24\textwidth}
		\includegraphics[width=\linewidth]{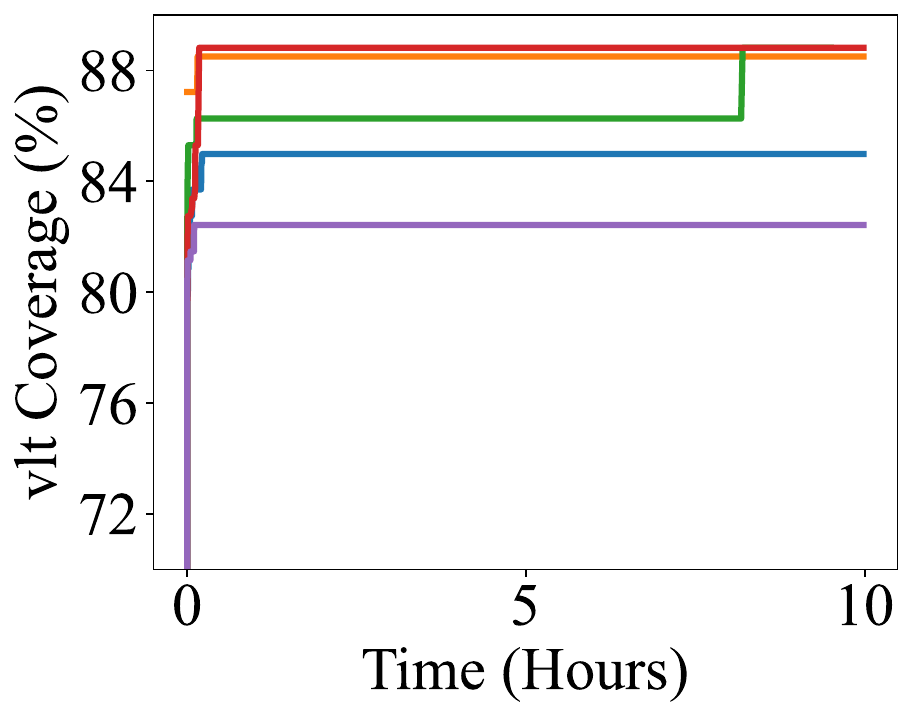}
		\caption{rv\_timer | HW Line (vlt)}
		\label{rv_HW_Line}
	\end{subfigure}

	\begin{subfigure}{\textwidth} 
		\centering
		\vspace{0.2cm}  
		\includegraphics[width=0.5\linewidth]{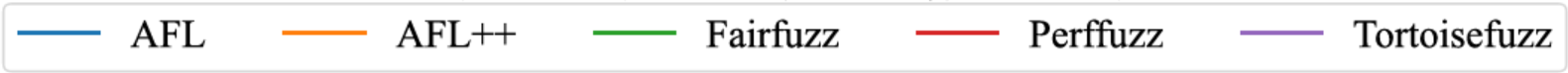} 
	\end{subfigure}
	
	\caption{Progression of hardware line coverage (vlt) during the application of FuzzWiz using various fuzzing engines on OpenTitan IP cores}
	\label{results}
\end{figure*}

%% file: sections/results.tex
With our \textit{FuzzWiz} framework, we were able to fuzz any hardware design written in most widely used \acp{HDL} such as VHDL, Verilog, and SystemVerilog. We evaluated this framework on multiple proprietary designs along with open-source ones.

\subsection{Designs under verification}
To show the significance and compatibility of \textit{FuzzWiz}, four \ac{IP} cores from Google's OpenTitan \cite{opentitan} project were selected as \acp{DUV} and are given below.

\begin{itemize}
	\small
	\item \ac{AES}
	\item \ac{KMAC}
	\item \ac{HMAC}
	\item \ac{RV-Timer}
\end{itemize}

These IPs play critical roles in security and system timing functions of the OpenTitan chip. \ac{AES} is crucial for data encryption and offers high-level security for various block sizes. \ac{KMAC} extends this security to message authentication and leverages Keccak sponge construction for adaptability. \ac{HMAC} uses SHA-256 as a hash function, and is crucial for maintaining data integrity in communication by authenticating messages. Lastly, \ac{RV-Timer} maintains system integrity with precise timing in RISC-V architectures. These cores were chosen because they represent fundamental components in cryptographic operations, system control, and longer time to reach coverage targets making them vital for benchmarking the efficacy of various fuzzing engines and their mutation strategies.

\subsection{Results and discussion}

In an industrial setup, design verification engineers run simulation regressions with random seeds over a certain period of time, for instance around 10 hours. After the regression completes, the generated results such as coverage metrics are further analyzed. Hence, we fuzzed the previously discussed \acp{DUV} multiple times for 10 hours to get meaningful results. The graphs represented in Fig. \ref{results} are generated using the best coverage results produced by each of those engines. From these results, it is evident that \textit{Fairfuzz} attained the highest hardware line coverage of around \SI{90}{\percent} for two out of four design IPs i.e., HMAC and RV-Timer as shown in Fig. \ref{hmac_HW_Line} and Fig. \ref{rv_HW_Line} respectively. This is primarily due to the input mutation strategy used by \textit{Fairfuzz}, with which it automatically prioritizes inputs exercising rare parts of the program under verification. But for AES and KMAC designs \textit{AFL++} fuzzer was predominant to reach a better coverage than \textit{Fairfuzz} in the same given time of 10 hours. This could be observed in Fig. \ref{aes_HW_Line} and Fig. \ref{kmac_HW_Line}. 

As the fuzzing progresses, it is observed from all the graphs shown in Fig. \ref{results} that the \textit{AFL++} fuzzer attains the coverage faster compared to other fuzzers because of its superior instrumentation and integration into Linux kernel module which improved its speed during the fuzzing process. In the course of our experiments, we observed that \textit{AFL++} has very little variance on the final coverage results achieved after fuzzing the \acp{DUV} multiple times with random seeds. Among all the fuzzers we applied, \textit{Perffuzz} and \textit{Tortoisefuzz} achieved comparatively lower coverage in the majority of the designs. These observations reveal that choosing an appropriate fuzzing engine is critical. Additionally, \ac{RTL} simulations of these IPs take 10 times more runtime than the fuzzing process to reach similar coverage \cite{opentitan_dashboard} \cite{google_interns}. To show the significance of our framework \textit{FuzzWiz}, we have compared it with other prior works as shown in Table \ref{comparison_works}. In contrast to previous works, \textit{FuzzWiz} is design-agnostic and comes with an option to choose which fuzzing engine is to be used. It also provides the testbenches with particular testvector (if there are any crashes during fuzzing process) to enhance the \ac{RTL} debugging capabilities. 

\begin{table}[htbp]
	\setcellgapes{1pt}
	\makegapedcells
	\renewcommand{\arraystretch}{1.3}
	\caption{Comparison of our work with other relevant works}
	\label{comparison_works}
	\footnotesize
	\centering
	\resizebox{0.5\textwidth}{!}{
		\begin{tabular}{|c|l|l|l|l|l|}
			\hline
			\textbf{Work}    & \textbf{\begin{tabular}[c]{@{}l@{}}Supported \\ HDL(s)\end{tabular}} & \textbf{Coverage Metric}                                              & \textbf{\begin{tabular}[c]{@{}l@{}}Fuzzing Engine/\\ Fuzzer\end{tabular}} & \textbf{\begin{tabular}[c]{@{}l@{}}Simulations\\ on\end{tabular}} & \textbf{\begin{tabular}[c]{@{}l@{}}Target \\ design type\end{tabular}} \\ \hline
			\cite{8587711}       & FIRRTL                                                               & Mux toggle                                                            & JQF                                                                       & FPGA                                                              & SoC                                                                    \\ \hline
			\cite{google_interns}      & V, SV                                                                & Edge coverage                                                         & AFL                                                                       & SW                                                                & SoC                                                                    \\ \hline
			\cite{9586289}      & FIRRTL                                                               & Mux toggle                                                            & JQF                                                                       & SW                                                                & SoC                                                                    \\ \hline
			\cite{9519470}        & FIRRTL, V                                                            & Control register                                                      & DIFUZZRTL                                                                 & FPGA                                                              & CPU                                                                    \\ \hline
			\cite{10133714}       & VHDL, SV                                                             & CSR transition                                                        & ProcessorFuzz                                                             & SW                                                                & CPU                                                                    \\ \hline
			\cite{10137024}       & V, SV                                                                & Cost-function                                                         & Modified AFL                                                              & FPGA                                                              & SoC                                                                    \\ \hline
			\cite{10192176}       & -                                                                    & Hardware code                                                         & WTF                                                                       & SW                                                                & IP                                                                     \\ \hline
			\textbf{\textit{FuzzWiz}} & VHDL, V, SV                                                          & \begin{tabular}[c]{@{}l@{}}Edge coverage or\\ Software line\end{tabular} & Multiple Engines                                                          & SW                                                                & IP, SoC                                                                \\ \hline
		\end{tabular}
	}
	\begin{center}
		\vspace{1ex}
		\justifying
		\scriptsize  Notes: Verilog (V), SystemVerilog (SV), Software (SW), Field Programmable Gate Array (FPGA)
	\end{center}
	\vspace{-0.7cm}
\end{table}

%% file: sections/conclusion.tex
In this paper, we presented an automated framework \textit{FuzzWiz} that can be used to perform fuzzing on hardware designs in order to reach coverage goals faster. We have applied \textit{FuzzWiz} on different internal and open-source designs to prove that it is design-agnostic. Unlike prior research, \textit{FuzzWiz} is compatible with various open-source software fuzzing engines. In order to show the importance and interoperability of the framework, it was applied on four \acs{IPs} from Google's OpenTitan. We explained how hardware fuzzing achieved around \SI{90}{\percent} of coverage for two out of four \acp{DUV}. We emphasized on the importance of choosing the correct fuzzing engine to reach coverage targets faster. We benchmarked the fuzzing engines by running them for 10 hours multiple times. From the results, we observed that \textit{Fairfuzz} performed better in terms of reaching the highest hardware line coverage for the specified time. However, \textit{AFL++} often improves the coverage as the time progresses during the fuzzing process. 